\documentclass[aps,prl,twocolumn,notitlepage]{revtex4-2}

\usepackage{amsmath}
\usepackage{amssymb}
\usepackage{graphicx}
\usepackage{epstopdf}
\usepackage[colorlinks=true]{hyperref}

\begin{document}
\title{Orbital angular momentum and current-induced motion of a Skyrmion-textured domain wall in a ferromagnetic nanotube}

\date{\today}

\author{Seungho Lee}
\affiliation{Department of Physics, Korea Advanced Institute of Science and Technology, Daejeon 34141, Republic of Korea}
\author{Se Kwon Kim}
\email{sekwonkim@kaist.ac.kr}
\affiliation{Department of Physics, Korea Advanced Institute of Science and Technology, Daejeon 34141, Republic of Korea}

\begin{abstract}
We theoretically study the current-induced dynamics of a domain wall in a ferromagnetic nanotube by developing a theory for the orbital angular momentum of a domain wall and the current-induced torque on it. Specifically, a domain wall with nontrivial magnetization winding along the circumference is shown to possess finite orbital angular momentum, which is proportional to the product of its Skyrmion charge and position, and the current is shown to exert a torque changing the orbital angular momentum of the domain wall and thereby drives it. The current-induced torque is interpreted as the transfer of orbital angular momentum from electrons to the domain wall, which occurs due to the emergent magnetic field associated with the Skyrmion charge. Our results reveal a hitherto unrecognized utility of the orbital degree of freedom of magnetic solitons.
\end{abstract}

\maketitle

\emph{Introduction.}|Ordered magnets exhibit various types of solitons that are stable for topological reasons~\cite{KosevichPR1990}. A prototypical example is a domain wall, which is a magnetic texture interpolating two uniform domains. Its dynamics driven by a current has been studied intensively because of fundamental interest as well as technological applications such as domain-wall racetrack memory~\cite{ParkinScience2008}. For example, it is well known that a spin-polarized current can drive a domain wall via spin-transfer torque~\cite{BergerPRB1996, SlonczewskiJMMM1996, TataraPRL2004, ThiavilleEPL2005}. The motion is rooted in the conservation of spin angular momentum: The total spin angular momentum of the magnet depends on the position of the wall. Electrons flip their spins while traversing the domain wall and thereby transfer their spin angular momentum to the wall. This results in the change of the spin angular momentum of the magnet, which manifests as the domain-wall motion.

One emerging platform for domain-wall dynamics is a magnetic nanotube~\cite{*[][{, and references therein.}] HertelJPCM2016}, which has gained attention for magnetic dynamics in curved geometry~\cite{*[][{, and references therein.}] StreubelJPD2016}. A domain-wall motion in a ferromagnetic nanotube has been studied with biases such as a magnetic field~\cite{LanderosJAP2010, YanAPL2011, YanAPL2012} and spin-transfer torque~\cite{OtaloraJPCM2012}, where the conservation of spin angular momentum is often invoked to explain the domain-wall motion. Besides the internal spin degree of freedom, a nanotube has an additional degree of freedom concerning rotations about the cylindrical axis. The potential relation between this orbital degree of freedom and the domain-wall dynamics has not been investigated yet.

\begin{figure}
\includegraphics[width=1\columnwidth]{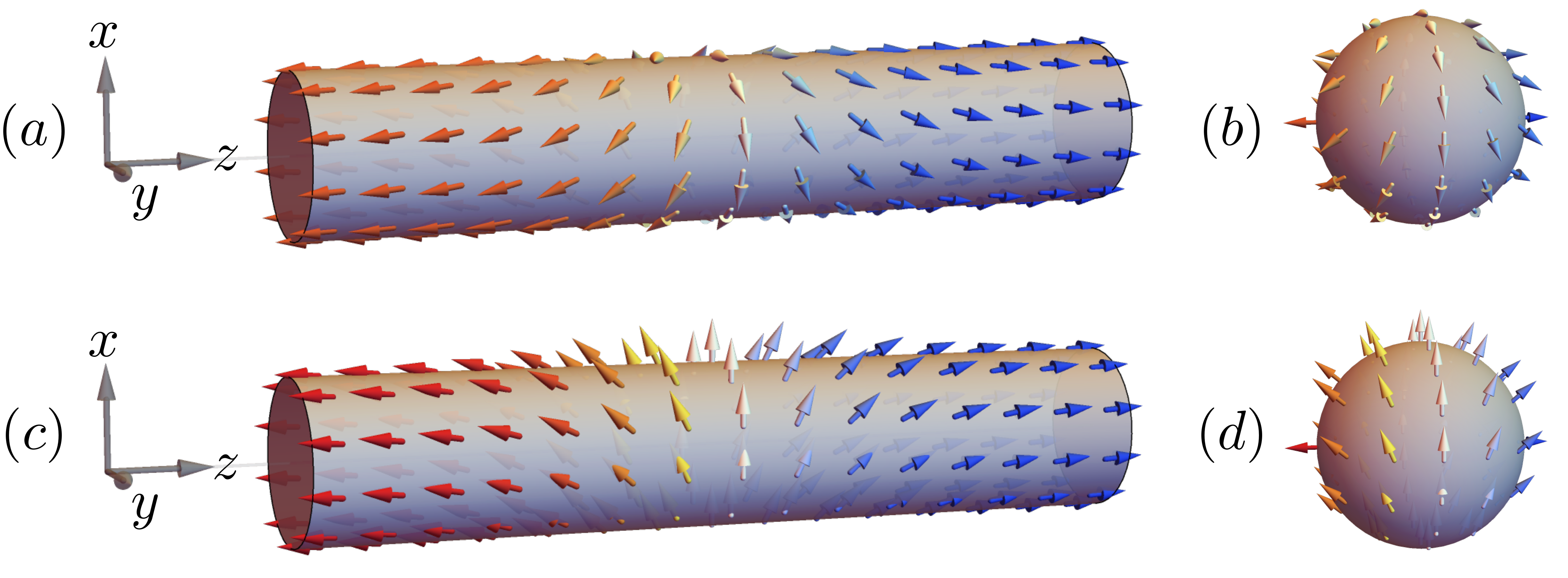}
\caption{Domain walls in a nanotube with (a) Skyrmion charge $Q = 1$ and (b) $Q = 0$, and their topologically equivalent configurations on the unit sphere (b) and (d), respectively.}
\label{fig:fig1}
\end{figure}

\emph{Main results.}|In this Letter, by accounting for the orbital degree of freedom of a magnetic nanotube, we develop a theory for the orbital angular momentum of a domain wall and its motion by the current-induced torque. To avoid potential confusion, we emphasize that the orbital angular momentum in this work refers to a physical quantity that is conserved when the nanotube has a global axial symmetry~\cite{YanPRB2013}, which is distinct from the locally-defined orbital angular momentum of electrons with respect to the crystal lattice. The linear momentum of a magnetic domain wall has been studied previously~\cite{HaldanePRL1986, VolovikJPC1987, YanPRB2013, TchernyshyovAP2015, DasguptaPRB2018}; our work addresses its orbital counterpart. 

Our model system is a thin ferromagnetic nanotube with easy-axis anisotropy in the $z$ direction, whose magnetic state is described by the unit vector $\mathbf{m}(z, \varphi)$, where $\varphi$ is the azimuthal angle. Let us consider a domain wall with a boundary condition $\mathbf{m} (z \rightarrow \pm \infty, \varphi) \rightarrow \pm \hat{\mathbf{z}}$. Since $\mathbf{m}$ converges to a single value as $z \rightarrow \pm \infty$, the magnetization profile is topologically equivalent to a certain unit vector field on the unit sphere. Then, like the latter, each domain wall is classified by the Skyrmion charge~\cite{SkyrmePRSA1961, BelavinJETP1975}:
\begin{equation}
\label{eq:Q}
Q = \frac{1}{4 \pi} \int dz d\varphi \, \mathbf{m} \cdot (\partial_\varphi \mathbf{m} \times \partial_z \mathbf{m}) = 0, \pm 1, \pm 2, \cdots \, ,
\end{equation}
which counts how many times $\mathbf{m}(z, \varphi)$ wraps the unit sphere as $z$ and $\varphi$ change. See Figs.~\ref{fig:fig1}(a) and (b) for a domain wall with Skyrmion charge $Q = 1$ and its topologically equivalent partner, respectively. Topologically trivial case ($Q = 0$) is depicted in Figs.~\ref{fig:fig1}(c) and (d). 

The dynamics of the magnetization in the presence of a current density $\mathbf{J} = J \hat{\mathbf{z}}$ is described by the Landau-Lifshitz equation~\cite{LL3}: $s \dot{\mathbf{m}} = \mathbf{h}_\text{eff} \times \mathbf{m} + P (\mathbf{J} \cdot \boldsymbol{\nabla}) \mathbf{m}$, where $s$ is the spin density, $\mathbf{h}_\text{eff} = - \delta U / \delta \mathbf{m}$ is the effective field, and $U$ is the potential energy. Here, the last term is the adiabatic spin-transfer torque~\cite{BergerPRB1996, SlonczewskiJMMM1996}, where $P = (\hbar/2e) \mathcal{P}$ and $\mathcal{P} = (\sigma_{\uparrow} - \sigma_{\downarrow})/(\sigma_{\uparrow} + \sigma_{\downarrow})$ is the dimensionless spin-polarization factor with spin-dependent conductivity $\sigma_s$ ($s = \uparrow$ or $s = \downarrow$ with $\uparrow$ chosen along $-\mathbf{m}$).

The low-energy dynamics of a domain wall can be described within the collective-coordinate approach~\cite{ThielePRL1973, TretiakovPRL2008, TretiakovPRL2012}. For our domain wall, in addition to the two conventional collective coordinates, position $Z(t)$ and spin-azimuthal angle $\Phi(t)$~\cite{TataraPRL2004, ThiavilleEPL2005}, we consider one more coordinate that is a domain-wall azimuthal angle $\Upsilon(t)$ associated with spatial rotations of the magnetization texture about the $z$ axis: $\mathbf{m}(z, \varphi, t) = \text{R}_z (\Phi (t)) \mathbf{m}_0 (z - Z (t), \varphi - \Upsilon(t))$, where $\text{R}_z (\psi)$ is the $3 \times 3$ rotation matrix about the $z$ axis by angle $\psi$. When the system is invariant under translations along the $z$ axis, spatial rotations about the $z$ axis, and spin rotations about the $z$ axis, the following momenta are conserved respectively: linear momentum, orbital angular momentum, and spin angular momentum. A domain wall breaks all the symmetries spontaneously and $Z$, $\Upsilon$, and $\Phi$ represent the associated zero-energy modes.

Within the collective-coordinate approach, the time evolution of the magnetization is given by $\dot{\mathbf{m}} = \dot{Z} \partial_Z \mathbf{m} + \dot{\Phi} \partial_\Phi \mathbf{m} + \dot{\Upsilon} \partial_\Upsilon \mathbf{m}$. By replacing the former by the latter in the Landau-Lifshitz equation, taking the inner product of both sides with $\mathbf{m} \times \partial_\varphi \mathbf{m}$, and integrating the resultant equation over the sample volume~\cite{ThielePRL1973, TretiakovPRL2008, TretiakovPRL2012}, we obtain the following equation of motion:
\begin{equation}
\label{eq:main}
- 4 \pi \rho s Q \dot{Z} = - \frac{\partial U}{\partial \Upsilon} + 4 \pi \rho Q P J \, ,
\end{equation}
where $\rho$ is the radius of the nanotube. Let us neglect the current for a moment by setting $J = 0$. The right-hand side $F_\Upsilon = -\partial U / \partial \Upsilon$ is a generalized force conjugate to the collective coordinate $\Upsilon$ associated with axial rotations, i.e., a torque~\cite{Goldstein} on the domain wall, which would be finite if the axial symmetry is broken. When the system respects the axial symmetry, the torque is zero and thus the left-hand side should vanish too. The conserved quantity associated with the rotational symmetry is orbital angular momentum, and thus we identify
\begin{equation}
\label{eq:L}
L_z = - 4 \pi \rho s Q Z \, ,
\end{equation}
as the orbital angular momentum of the domain wall, which is our first main result. It is a linear function of the wall position $Z$ similarly to the spin angular momentum of the domain wall. Also, note that the orbital angular momentum is proportional to its Skyrmion charge. 

We present an alternative derivation of Eq.~(\ref{eq:L}) within the Lagrangian formalism based on the Lagrangian $L=\rho s \int dz d\varphi \mathbf{a}(\mathbf{m})\cdot \dot{\mathbf{m}}-U$, where $\mathbf{a}(\mathbf{m})$ is the vector potential for the monopole implicitly defined by $\boldsymbol{\nabla}_\mathbf{m} \times \mathbf{a} = \mathbf{m}$~\cite{Altland}. By invoking Noether's theorem~\cite{Goldstein} with the Lagrangian, \textcite{YanPRB2013} has derived the expression for the density of the orbital angular momentum of axially symmetric ferromagnets, and it is given by $\mathcal{L}_z = s \cos \theta \partial_\varphi \phi$. The orbital angular momentum $L_z$ of a domain wall is given by the spatial integration of its density. By exploiting the azimuthal periodicity of the density $\mathcal{L}_z (z, \varphi + 2 \pi) = \mathcal{L}_z (z, \varphi)$, one can show that $\partial_\Phi L_z = 0$, $\partial_\Upsilon L_z=0$, and $\partial_Z L_z = -\rho s\int dz d\varphi \sin \theta (\partial_\varphi \theta \partial_z \phi - \partial_\varphi \phi \partial_z \theta) =-4\pi\rho sQ$. Note that $L_z$ depends only on the domain-wall position $Z$. These derivatives of $L_Z$ indicate $L_Z = -4\pi\rho sQ Z$, which is identical to the result Eq.~(\ref{eq:L}) obtained from the Landau-Lifshitz equation (up to an irrelevant constant). 

\begin{figure}
\includegraphics[width=0.9\columnwidth]{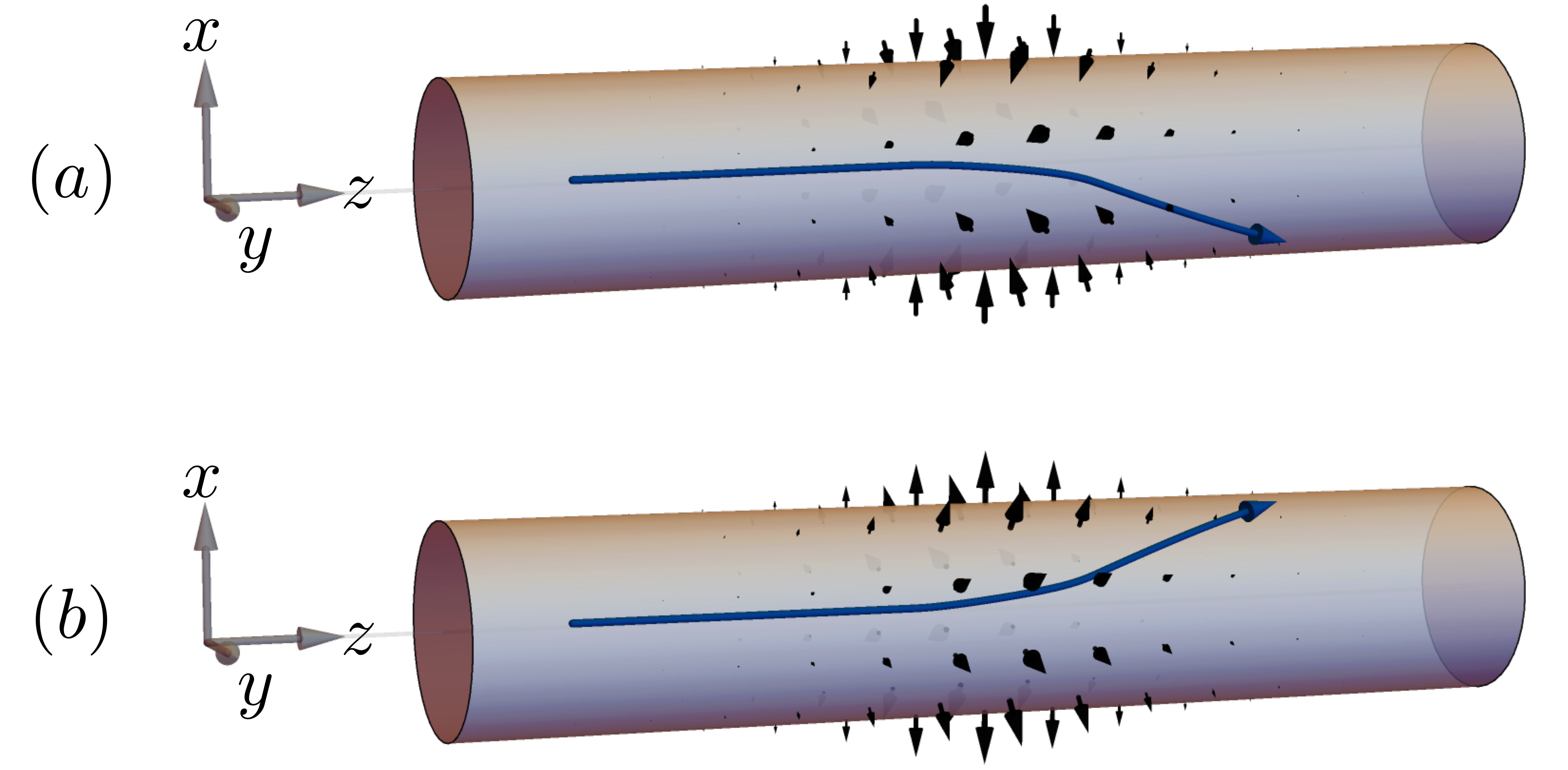}
\caption{Schematic illustration of the emergent magnetic field $\mathbf{b}$ [Eq.~(\ref{eq:b})] (denoted by black arrows) of domain walls with (a) $Q = 1$ and (b) $Q = -1$ and the corresponding electron trajectories (shown as blue arrows).}
\label{fig:fig2}
\end{figure}

Let us now consider the effect of a current, $J \neq 0$. Since the left-hand side of Eq.~(\ref{eq:main}) is the time evolution of the domain-wall orbital angular momentum, we identify
\begin{equation}
\label{eq:ott}
F_\Upsilon^J = 4 \pi \rho Q P J \, ,
\end{equation}
as a torque on the domain wall, which, we note, is proportional to the Skyrmion charge of the domain wall. The adiabatic spin-transfer torque describes the effect of an electron current on a magnetization texture when electrons keep their spins collinear with the local magnetization. The physical interpretation of the torque [Eq.~(\ref{eq:ott})] can be obtained by invoking its reciprocal effect of a magnetization texture on an electron current~\cite{VolovikJPC1987, ZangPRL2011, NagaosaPS2012, NagaosaNN2013}. When electrons (whose spin follows the direction of the local spin density) traverse the magnetization texture, they experience the emergent magnetic field $b_i = (\hbar/4q) \epsilon_{ijk} \mathbf{m} \cdot (\partial_j \mathbf{m} \times \partial_k \mathbf{m})$, where $q = - e < 0$ is the charge of electrons~\cite{VolovikJPC1987, ZangPRL2011, NagaosaPS2012, NagaosaNN2013}, which has been demonstrated as the topological Hall effect in skyrmion-crystal phases of chiral magnets~\cite{LeePRL2009, NeubauerPRL2009, HuangPRL2012the, LiPRL2013the}. For a domain wall in a nanotube, it is in the radial direction $\mathbf{b} = b \hat{\mathbf{e}}_\rho$ with
\begin{equation}
\label{eq:b}
b = \frac{\hbar}{2 q \rho} \mathbf{m}_0 \cdot (\partial_\varphi \mathbf{m}_0 \times \partial_z \mathbf{m}_0) \, ,
\end{equation}
which is nothing but the Skyrmion-charge density besides the constant factor [Eq.~(\ref{eq:Q})]. In the presence of the radial magnetic field, an electron moving on the nanotube experiences the corresponding Lorentz force, and thus its azimuthal velocity evolves by $m \dot{v}_\varphi = q \dot{z}(t) b(z(t), \varphi(t))$, where $m$ is the effective mass of the electron and $(z(t), \varphi(t))$ represent its position. Then, the change of the azimuthal velocity after the electron passes through the wall is given by $\Delta v_\varphi = (q/m) \int \, b(z, \varphi) (dz/dt) dt = (q/m) \int b(z, \varphi) dz$. By taking the average of this result over the azimuthal angle $\varphi$, the average change of the orbital angular momentum of the electron is given by
\begin{equation}
\label{eq:lz}
\langle \Delta l_z \rangle = \rho m \langle \Delta v_\varphi \rangle = Q \hbar \, .
\end{equation}
This indicates that each electron (whose spin is antiparallel to the local magnetization) passing the domain wall should transfer $- Q \hbar$ to the domain wall. After accounting for partial spin polarization, the rate of the transfer of orbital angular momentum from the current to the domain wall is given by $Q \hbar \times (2 \pi \rho \mathcal{P} J) / e = 4 \pi \rho Q P J$. This is identical to the torque [Eq.~(\ref{eq:ott})] derived from the adiabatic spin-transfer torque. See Figs.~\ref{fig:fig2}(a) and~\ref{fig:fig2}(b) for the emergent magnetic field $\mathbf{b}$ and the corresponding electron trajectory when $Q = 1$ and $Q = -1$, respectively. The torque [Eq.~(\ref{eq:ott})] induced by the adiabatic spin-transfer torque and its interpretation based on the reciprocal effect, namely the emergent magnetic field [Eq.~(\ref{eq:b})] and the resultant change of electron's orbital angular momentum [Eq.~(\ref{eq:lz})], constitute our second main result. We remark that our torque originates from the isotropic exchange interaction between local magnetization and electron's spin with no regard to spin-orbit coupling and is thus distinct from the orbital torque studied in Refs.~\cite{GoPRL2018, GoPRR2020-2}. 

Equation~(\ref{eq:main}) indicates that the torque can drive a domain wall at velocity $\dot{Z} = - P J / s$ when the system respects the axial symmetry, i.e., $\partial U / \partial \Upsilon = 0$. This mechanism works for the domain wall with finite Skyrmion charge $Q \neq 0$ whenever the axial symmetry is present in the system. Below, by using an explicit solution, we will further discuss the orbital angular momentum of a domain wall and its motion by the current-induced torque.

\emph{Explicit model.}|We consider a ferromagnetic nanotube whose potential-energy density is given by
\begin{equation}
\label{eq:U}
\mathcal{U} = \frac{A}{2} \left[ \left( \partial_z \mathbf{m} \right)^2 + \frac{\left( \partial_\varphi \mathbf{m} \right)^2 }{\rho^2} \right] - \frac{K }{2} m_z^2 + \frac{\kappa}{2} (\mathbf{m} \cdot \hat{\mathbf{e}}_\rho)^2 \, ,
\end{equation}
where $A$ is the exchange coefficient and $K > 0$ is the easy-axis anisotropy. Here, $\kappa > 0$ represents the effect of the dipolar interaction that favors the magnetization lying on the surface~\cite{LanderosJAP2010}, and it is linearly proportional to the thickness of the nanotube~\cite{LanderosJAP2010, MakhfudzPRL2012}. We consider the limit of thin nanotubes by setting $\kappa = 0$, by focusing on the cases where an applied current is large enough to dominate the effect of the dipolar interaction~\footnote{See the Supplemental Material for the discussions of certain cases with finite $\kappa$ and a domain-wall solution in the presence of an external field.}. 

In terms of spherical angles, the magnetization is given by $\mathbf{m} = (\sin \theta \cos \phi, \sin \theta \sin \phi, \cos \theta)$. A domain-wall solution is a stationary state with a boundary condition $\mathbf{m}(z \rightarrow \pm \infty) = \pm \hat{\mathbf{z}}$. A family of solutions is given by
\begin{equation}
\label{eq:dw}
\begin{aligned}
\cos \theta_0 &= \tanh \left[ \frac{(z - Z) \sqrt{1 + n^2 \lambda_0^2/\rho^2}}{\lambda_0} \right] \, , \\
\phi_0 &= n (\varphi - \Upsilon) + \Phi \, , \quad n = 0, \pm 1, \pm 2, \cdots \, ,
\end{aligned}
\end{equation}
where $\lambda_0 = \sqrt{A/K}$ and $n$ is a winding number that counts how many times that the magnetization winds the unit circle in the $xy$ plane as $\varphi$ changes from $0$ to $2 \pi$. In literature, $n = 0$ and $n = 1$ solutions are referred to as transverse and vortex-like domain walls~\cite{*[][{, and references therein.}] HertelJPCM2016}, respectively. Note that the domain-wall width is given by $\lambda_n = \lambda_0 / \sqrt{1 + n^2 \lambda_0^2/\rho^2}$. The Skyrmion charge [Eq.~(\ref{eq:Q})] of the domain wall equals the winding number: $Q = n$. 

The equations of motion for the three collective coordinates obtained with the explicit domain-wall solution from the Landau-Lifshitz equation~\cite{ThielePRL1973, TretiakovPRL2008} are given by
\begin{eqnarray}
- 4 \pi \rho s \dot{\Phi} + 4 \pi \rho s n \dot{\Upsilon} &=& F_Z^J = 0 \, , \label{eq:P1} \\
4 \pi \rho s \dot{Z} &=& F_\Phi^J = - 4 \pi \rho P J \, , \label{eq:S1} \\
- 4 \pi \rho s n \dot{Z} &=& F_\Upsilon^J = 4 \pi \rho n P J \, , \label{eq:L1} 
\end{eqnarray}
where the last equation is identical to Eq.~(\ref{eq:main}) with $Q = n$.

The right-hand side of Eq.~(\ref{eq:P1}) is the current-induced force on the domain wall, which can be engendered by, e.g., nonadiabatic spin-transfer torque~\cite{ThiavilleEPL2005, TserkovnyakJMMM2008}. The left-hand side is the time derivative of the linear momentum:
$P_z = - 4 \pi \rho s (\Phi - Q \Upsilon) \, ,$
which has been derived previously~\cite{HaldanePRL1986, VolovikJPC1987, YanPRB2013, TchernyshyovAP2015, DasguptaPRB2018} for $Q = 0$. The left-hand side of Eq.~(\ref{eq:S1}) is the time derivative of the spin angular momentum of the domain wall, which is given by $S_z = 4 \pi \rho s Z$. Alternatively, this can be obtained directly by integrating the spin density, $- \rho s \int dz d\varphi (\mathbf{m}_0 \cdot \hat{\mathbf{z}})$ over the volume. The right-hand side of Eq.~(\ref{eq:S1}) is the spin torque on the domain wall, whose physical mechanism can be understood as follows. When electrons, whose spin is kept antiparallel to $\mathbf{m}$, traverse the domain wall, their spins change from $(\hbar/2) \hat{\mathbf{z}}$ to $- (\hbar/2) \hat{\mathbf{z}}$ (assuming $J < 0$). Consequently, the domain wall absorbs $\hbar \hat{\mathbf{z}}$ per one electron. After accounting for partial spin polarization, the total spin angular momentum transfer from electrons to the domain wall per unit time is given by $(\hbar/e) \rho \mathcal{P}|J| \times 2 \pi = - 4 \pi \rho P J$, the right-hand side of Eq.~(\ref{eq:S1}).

\emph{Current-driven dynamics.}|The significance of the orbital angular momentum of a domain wall manifests clearly when the spin-rotational symmetry is broken so that the spin angular momentum is not conserved, but orbital angular momentum is. As an illustrative case, let us consider an external field in the $x$ direction $\mathbf{H} = H \hat{\mathbf{x}}$. Also, we restore the damping $\alpha > 0$ hereafter, which gives rise to the viscous terms in the equations of motion~\cite{Note1}. The external field alters the domain-wall configuration from Eq.~(\ref{eq:dw}). In particular, the spin azimuthal angle is given by $\phi(z, \varphi) = f(z - Z, \varphi - \Upsilon) + \Phi$ with a certain nonlinear function $f$ differing from Eq.~(\ref{eq:dw})~\cite{Note1}. The function $f$ can be chosen such that $\Phi = 0$ corresponds to the lowest Zeeman energy, in which the domain-wall energy is given by $U(Z, \Phi, \Upsilon) = - k \cos (\Phi)$ where $k \equiv M H \int dz d\varphi \sin\theta(z, \varphi) \cos f(z, \varphi)$ is a positive coefficient and $M$ is the magnetization. The energy dependence on $\Phi$ indicates that $\Phi$ is no longer a zero-energy mode, while $Z$ and $\Upsilon$ remain as zero-energy modes. 

\begin{figure}
\includegraphics[width=0.9\columnwidth]{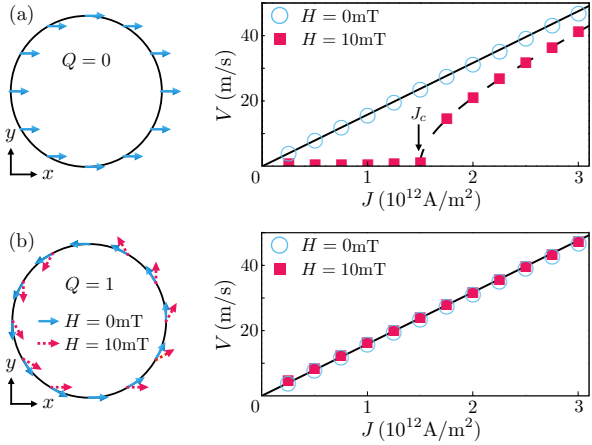}
\caption{The magnetization along the circumference and the domain-wall speed $V=\dot{\left\vert\langle {Z} \rangle\right\vert}$ as a function of a current density $J$ for domain walls with Skyrmion charge (a) $Q=0$ and (b) $Q=1$. Filled squares (open circles) represent simulations results in the absence (presence) of the external fields $\mathbf{H} = 10$mT$\hat{\mathbf{x}}$. Lines depict analytical results, Eq.~(\ref{eq:V0}) and Eq.~(\ref{eq:V1}).}
\label{fig:fig3}
\end{figure}

In the presence of the external field, the equations of motion for the collective coordinates are given by~\footnote{By treating the external field as a perturbation, we neglect the field-induced change of the gyrotropic and the dissipation coefficients in the equations of motion.}
\begin{eqnarray}
- s \lambda_n \dot{\Phi} + s \lambda_n Q \dot{\Upsilon} + \alpha s \dot{Z} &=& 0 \, , \label{eq:Pz} \\
s \dot{Z} + \alpha s \lambda_n \dot{\Phi} - \alpha s \lambda_n Q \dot{\Upsilon} &=& \frac{k \sin \Phi}{4 \pi \rho}  - P J \, , \\
- s Q \dot{Z} + \alpha s \lambda_n Q^2 \dot{\Upsilon} - \alpha s \lambda_n Q \dot{\Phi} &=& Q P J \, . \label{eq:Lz}
\end{eqnarray}
First, for $Q = 0$ case, the energy of the domain wall is described with $k = 2 \pi^2 \rho \lambda_0 M H$ and Eq.~(\ref{eq:Lz}) vanishes identically. The resultant average velocity $ \langle \dot{Z} \rangle$ is that of the well-known case of a one-dimensional motion driven by the adiabatic spin-transfer torque~\cite{TataraPRL2004, ThiavilleEPL2005}:
\begin{equation}
\label{eq:V0}
\langle \dot{Z} \rangle = - \frac{P \sqrt{J^2 - J_c^2}}{(1 + \alpha^2) s} \Theta(J - J_c) \, , \quad \text{for } Q = 0 \, ,
\end{equation}
where $\Theta(x)$ is the Heaviside step function and $J_c \equiv \pi \lambda_0 M H / 2 P$ is the critical current for the so-called Walker breakdown~\cite{SchryerJAP1974}. When the current $J$ (assumed positive without loss of generality) is smaller than the critical current $J < J_c$, the domain wall does not move $\dot{Z} = 0$ and the magnetization angle is kept constant $\Phi = \arcsin(2 P J / \pi \lambda_0 H)$. This shows that, when the current effect is weaker than the effect of the spin-symmetry-breaking field, adiabatic spin-transfer torque cannot drive a domain wall. When $J > J_c$, the spin-transfer torque drives the domain wall at a finite velocity.

Let us now turn to $Q \neq 0$ cases. Equations~(\ref{eq:Pz}-\ref{eq:Lz}) possess, regardless of the current magnitude, a steady-state solution with the constant velocity:
\begin{equation}
\label{eq:V1}
\langle \dot{Z} \rangle = - \frac{P J}{(1 + \alpha^2) s} \, , \quad \text{for } Q \neq 0 \, ,
\end{equation}
with constant spin azimuthal angle $\Phi = 0$ and constant angular velocity given by $\dot{\Upsilon} = (\alpha P J)/\{(1 + \alpha^2) s n \lambda_n\}$. The domain wall moves by rotating the Skyrmion texture around the circumference while keeping the finite magnetization in the $x$ direction constant. In this way, even in the presence of an external field that breaks the spin-rotational symmetry, a domain wall with $Q \neq 0$ can be driven by an arbitrarily small current. 

To confirm our analytical results, we performed micromagnetic simulations by solving the discrete Landau-Lifshitz-Gilbert equations with the energy $U$ [Eq.~(\ref{eq:U})] (with $\kappa = 0$) with the aid of $\textsc{MuMax}^3$~\cite{VansteenkisteAA2014}. The considered geometry is 3072nm$\times$96nm$\times$3nm (length$\times$circumference$\times$thickness) with lattice constant 3nm. We used typical material parameters of Co: $M_s=1.44\text{MA/m},\,A=31\text{pJ/m},\,K=410\text{kJ/}\text{m}^3,\,\mathcal{P}=0.4$~\cite{Coey} and $\alpha=0.1$. The simulation time for each case was 30ns~\footnote{For $J=1.75\times10^{12}\text{A/}\text{m}^2,\,Q=0,\,H=10\text{mT}$, the simulation time was 50ns to obtain a properly averaged velocity}. The magnetization textures, the simulation results, and the analytical results for $Q$=0 and $Q$=1 are shown in Figs.~\ref{fig:fig3}(a) and (b), respectively. In the left panels, the magnetization along the circumference in the absence of an external field is depicted schematically as blue arrows. When the field $\mathbf{H} = H \hat{\mathbf{x}}$ is applied, the configuration changes for the $Q = 1$ domain wall as shown as red arrows. In the right plots, filled squares (open circles) represent the simulation results for $H = 10$mT ($H = 0$mT). Lines show the analytical results, Eq.~(\ref{eq:V0}) for $Q = 0$ and Eq.~(\ref{eq:V1}) for $Q = 1$, which agree with the simulation results well. Note that, in the presence of an external field, a domain wall with $Q = 0$ does not move below the critical current $J_c \approx 1.5 \times 10^{12}$A/m$^2$, whereas a domain wall with $Q = 1$ moves regardless of the current magnitude, as expected from our analysis.

\emph{Summary.}|We have developed a theory for the orbital angular momentum of a domain wall in a ferromagnetic nanotube and the current-induced torque on it, which we have interpreted as the transfer of orbital angular momentum of electrons to the domain wall caused by the emergent magnetic field associated with the Skyrmion charge of the wall. We have shown theoretically and numerically that the orbital degree of freedom of magnetic nanotubes engenders a transfer channel of orbital angular momentum between a domain wall and electrons, through which the wall with finite Skyrmion charge can be driven. We hope that our results on the orbital degree of freedom of domain walls are extended to other solitons such as Skyrmions and vortices.

\begin{acknowledgments}
We are grateful to Kyung-Jin Lee, Kouki Nakata, Pengtao Shen, and Oleg Tchernyshyov for insightful discussions. This work is supported by Brain Pool Plus Program through the National Research Foundation of Korea funded by the Ministry of Science and ICT (Grant No. NRF-2020H1D3A2A03099291) and by the National Research Foundation of Korea funded by the Korea Government via the SRC Center for Quantum Coherence in Condensed Matter (Grant No. NRF-2016R1A5A1008184).
\end{acknowledgments}

\bibliography{/Users/kimsek/Dropbox/School/Research/master}

\pagebreak
\appendix

\setcounter{equation}{0}
\setcounter{figure}{0}
\setcounter{table}{0}
\setcounter{page}{1}
\renewcommand{\theequation}{S\arabic{equation}}
\renewcommand{\thefigure}{S\arabic{figure}}

\onecolumngrid
\begin{center}
\textbf{\large Supplemental Material: Orbital angular momentum of a magnetic domain wall and the motion by orbital-transfer torque} \\[10pt]
Seungho Lee and Se Kwon Kim\\
\textit{Department of Physics, Korea Advanced Institute of Science and Technology, Daejeon 34141, Republic of Korea}
\end{center}
\twocolumngrid

\section{1. Collective-coordinate approach}
\label{sec:app1}

The dynamics of a domain wall can be described within the collective-coordinate formalism~\cite{ThielePRL1973, TretiakovPRL2008}. The Landau-Lifshitz-Gilbert equation with the adiabatic spin-transfer torque is given by
\begin{equation}
s \dot{\mathbf{m}} - \alpha s \mathbf{m} \times \dot{\mathbf{m}} = \mathbf{h}_\text{eff} \times \mathbf{m} + P (\mathbf{J} \cdot \boldsymbol{\nabla}) \mathbf{m} \, , 
\end{equation}
where
\begin{equation}
\mathbf{h}_\text{eff} \equiv - \frac{\delta U}{\delta \mathbf{m}} \, ,
\end{equation}
is the effective field. In the collective-coordinate approach, we assume that the time evolution of the magnetization $\mathbf{m}(\mathbf{r}, t)$ can be captured by the time evolution of a few collective coordinates $\xi_i (t)$: $\mathbf{m}(\mathbf{r}, t) = \mathbf{m}_0 (\mathbf{r}, \boldsymbol{\xi}(t))$. Within this assumption, the time derivative of the magnetization can be written as follows:
\begin{equation}
\dot{\mathbf{m}} = \dot{\xi}_j \frac{\partial \mathbf{m}_0}{\partial \xi_j} \, ,
\end{equation}
where the Einstein summation over the index $i$ is assumed. Upon replacing $\dot{\mathbf{m}}$ by the above expression, the LLG equation becomes
\begin{equation}
\dot{\xi}_j \frac{\partial \mathbf{m}_0}{\partial \xi_j} - \alpha s \dot{\xi}_j \mathbf{m}_0 \times \frac{\partial \mathbf{m}_0}{\partial \xi_j} = \mathbf{h}_\text{eff} [\mathbf{m}_0] \times \mathbf{m}_0 + P (\mathbf{J} \cdot \boldsymbol{\nabla}) \mathbf{m}_0 \, .
\end{equation}
After taking the inner product between both sides and $\mathbf{m}_0 \times (\partial \mathbf{m}_0 / \partial \xi_j)$ followed by the integration over the volume, we obtain
\begin{equation}
- G_{ij} \dot{\xi}_j + D_{ij} \dot{\xi}_j = F_i + F_i^J \, ,
\end{equation}
where 
\begin{eqnarray}
G_{ij} &\equiv& s \int dV \mathbf{m}_0 \cdot \left( \frac{\partial \mathbf{m}_0}{\partial \xi_i} \times \frac{\partial \mathbf{m}_0}{\partial \xi_j} \right) \, , \\
D_{ij} &\equiv& \alpha s \int dV \left( \frac{\partial \mathbf{m}_0}{\partial \xi_i} \cdot \frac{\partial \mathbf{m}_0}{\partial \xi_j} \right) \, , \\
F_i &\equiv& - \frac{d U}{d \xi_i} \, , \\
F_i^J &\equiv& - P \int dV \, \mathbf{m}_0 \cdot \left( \frac{\partial \mathbf{m}_0}{\partial \xi_j} \times (\mathbf{J} \cdot \boldsymbol{\nabla}) \mathbf{m}_0 \right) \, ,
\end{eqnarray}
are called the gyrotropic tensor, the dissipation tensor, and the equilibrium force, and the current-induced force, respectively. The gyrotropic tensor and the dissipation tensor are antisymmetric and symmetric with respect to the exchange of indices, respectively. This formalism can be applied to dynamics of various type of magnetic textures including domain walls, vortices, and Skyrmions.

For the domain wall [Eq.~(\ref{eq:dw})], these are given by
\begin{eqnarray}
G_{Z \Phi} &=& 4 \pi \rho s \, , \\
G_{Z \Upsilon} &=& - 4 \pi \rho s Q \, , \\
G_{\Phi \Upsilon} &=& 0 \, , \\
D_{Z Z} &=& 4 \pi \alpha \rho s / \lambda_n \, , \\
D_{Z \Phi} &=& 0 \, , \\
D_{Z \Upsilon} &=& 0 \, , \\
D_{\Phi \Phi} &=& 4 \pi \alpha \rho s \lambda_n \, , \\
D_{\Phi \Upsilon} &=& - 4 \pi \alpha \rho s n \lambda_n \, , \\
D_{\Upsilon \Upsilon} &=& 4 \pi \alpha \rho s n^2 \lambda_n \, , \\
F_Z^J &=& - 4 \pi n P J_\varphi \, , \\
F_\Phi^J &=& - 4 \pi \rho P J_z  \, , \\
F_\Upsilon^J &=& 4 \pi \rho Q P J_z \, ,
\end{eqnarray}
where the skyrmion charge $Q$ is equal to the magnetization winding number $n$. Here, the expressions for the two quantities related to $\Upsilon$, $G_{Z \Upsilon} = - 4 \pi \rho s Q$ and $F_\Upsilon^J = 4 \pi \rho Q P J_z$, can be obtained without the explicit domain-wall solution as shown in the main text. Also, $G_{\Phi \Upsilon} = 0$ can be derived without the explicit solution as follows: $G_{\Phi \Upsilon} = s \int dz d\varphi \mathbf{m}_0 \cdot (\partial_\Phi \mathbf{m}_0 \times \partial_\Upsilon \mathbf{m}_0) = s \int dz d\varphi \sin \theta \partial_\varphi \theta = 0$.

\section{2. A domain wall in the presence of an external field $\mathbf{H} = H \hat{\mathbf{x}}$}
\label{sec:app2}

Here, we discuss a domain-wall solution in the presence of the Zeeman energy, $U_Z = - \rho M H \int dz d\varphi \, \sin \theta \cos \phi$. The solution given in Eq.~(\ref{eq:dw}) is altered by the external field. The exact solution is unavailable and we use the following variational ansatz for the domain wall:
\begin{equation}
\begin{aligned}
\cos \theta_0 &= \tanh \left[ \frac{(z - Z) \sqrt{1 + n^2 \lambda_0^2/\rho^2}}{\lambda_0} \right] \, , \\
\phi_0 &= n (\varphi - \Upsilon) - c \sin (\varphi - \Upsilon) + \Phi \, ,
\end{aligned}
\end{equation}
and look for the value of the variational parameter $c$ that minimizes the energy. The energy terms that depend on the variational parameter $c$ is given by
\begin{eqnarray}
&& \rho \int dz d\varphi \left[ \frac{A \sin^2 \theta (\partial_\varphi \phi)^2}{2 \rho^2} - M H \sin \theta \cos \phi \right] \, , \\
&=& \frac{A \lambda_n}{\rho} \int d\varphi (n - c \sin \varphi)^2 \\
&& - \pi \rho \lambda_n M H \int d\varphi \cos (n \varphi - c \sin \varphi + \Phi) \, \\
&=& \frac{\pi \lambda_n A (2 n^2 + c^2)}{\rho} - 2 \pi^2 \rho \lambda_n M H J_n(c) \cos \Phi \, , 
\end{eqnarray}
where $J_n (x)$ is the Bessel function of the first kind. For $H > 0$, the minimum energy is obtained for $\Phi = 0$ and the variational parameter $c$ is implicitly given by $J'_n(c) / c = A / (\pi \rho^2 M H)$. In the limit of strong field $H \gg A / (\pi \rho^2 M)$, the Zeeman term dominates and the optimal value for the variation parameter is given by the first point where $J'_n (c) = 0$, which yields, e.g., $c = 0$ for $Q = 0$. The potential energy now depends on $\Phi$ as $U(Z, \Upsilon, \Phi) = - k \cos \Phi$ with $k = 2 \pi^2 \rho M H J_n(c)$ evaluated at the optimal value for $c$. For example, $k = 2 \pi^2 \rho \lambda_0 M H$ for $Q = 0$, which agrees with the result obtained from Eq.~(\ref{eq:dw}).

\section{3. The effect of the dipolar interaction (finite $\kappa$ case)}
\label{sec:app3}

Let us discuss the dynamics of domain walls in the case of $\kappa > 0$ which captures the effect of the dipolar interaction. In terms of the spherical angles $\mathbf{m} = (\sin \theta \cos \phi, \sin \theta \sin \phi, \cos \theta)$, the energy-density term $\propto \kappa$ is given by $(\kappa/2) \sin^2 \theta \cos^2 (\phi - \varphi)$. This term breaks each of the spin-rotational symmetry and the axial symmetry, but there is a residual symmetry. The energy is invariant under simultaneous rotation of spin and space: $\phi, \varphi \mapsto \phi + \Delta, \varphi + \Delta$.

We treat the term $\propto \kappa$ as a perturbation and use the previous domain-wall solution as an ansatz. Then, the energy is no longer independent of $\Phi$ and $\Upsilon$. Instead, we have
\begin{eqnarray}
U(\Phi, \Upsilon) &=& \rho \int dz d\varphi \left[ \frac{\kappa}{2} \sin^2 \theta_0 \cos^2 (\phi_0 - \varphi) \right] \, , \\
&=& \rho \kappa \int d\varphi \, \cos^2 ((n - 1) \varphi + \Phi - n \Upsilon) \, .
\end{eqnarray}

The lowest-energy state is given by $n = 1$ (and thus $Q = 1$) (which is a vortex domain-wall state studied in Ref.~\cite{HertelJPCM2016}). We focus on this case now. Then, the energy is given by
\begin{equation}
U(\Phi, \Upsilon) = 2 \pi \rho \kappa \cos^2 (\Phi - \Upsilon) \, .
\end{equation}

The equations of motion in the presence of the current density $\mathbf{J} = J \hat{\mathbf{z}}$ are given by
\begin{equation}
  - 4 \pi \rho s \dot{\Phi} + 4 \pi \rho s Q \dot{\Upsilon} + 4 \pi \alpha \rho s \dot{Z} / \lambda_n = 0 \, ,
\end{equation}
for the position $Z$, 
\begin{equation}
\begin{aligned}
& 4 \pi \rho s \dot{Z} + 4 \pi \alpha \rho s \lambda_n \dot{\Phi} - 4 \pi \alpha \rho s n \lambda_n \dot{\Upsilon} \\
& = - 4 \pi \rho P J + 4 \pi \rho \kappa \cos (\Phi - \Upsilon) \sin (\Phi - \Upsilon)  \, ,
\end{aligned}
\end{equation}
for the spin angle $\Phi$, and
\begin{equation}
\begin{aligned}
& - 4 \pi \rho s Q \dot{Z} + 4 \pi \alpha \rho s n^2 \lambda_n \dot{\Upsilon} - 4 \pi \alpha \rho s n \lambda_n \dot{\Phi} \\
& = 4 \pi \rho Q P J - 4 \pi \rho \kappa \cos (\Phi - \Upsilon) \sin (\Phi - \Upsilon) \, ,
\end{aligned}
\end{equation}
for the orbital angle $\Upsilon$. It is more convenient to use $(\Xi \equiv \Phi - Q \Upsilon, \Phi)$ instead of $(\Upsilon, \Phi)$. In terms of $\Xi$ and $\Phi$, we have
\begin{eqnarray}
&& \lambda_n \dot{\Xi} - \alpha \dot{Z} = 0 \, , \\
&& - s \dot{Z} - P J_z + \kappa \cos \Xi \sin \Xi - \alpha s \lambda_n \dot{\Xi} = 0 \, .
\end{eqnarray}

For a small current, the domain wall does not move $\dot{Z} = 0$. For the sufficiently large current so that $|J| \gg J_c \equiv \kappa / 2 P$ at which the Walker breakdown occurs~\cite{SchryerJAP1974}, we have
\begin{equation}
\dot{Z} \approx - \frac{P J}{(1 + \alpha^2) s} \, .
\end{equation}

\end{document}